\documentclass[aps,prl,twocolumn,floatfix]{revtex4}

\pdfoutput=1

\usepackage{amsmath}
\usepackage{amsfonts}
\usepackage{bm} % bold math
\usepackage{braket}
\usepackage{color}
\usepackage{colortbl}
\usepackage{dcolumn}
\usepackage{epstopdf}
\usepackage{float}
\usepackage{flushend}
\usepackage{graphicx}
\usepackage{xcolor}
\usepackage{subfigure}

\renewcommand{\gg}{\mathbf{g}}

\newcommand{\GG}{\mathbf{G}}

\newcommand{\kk}{\mathbf{k}}
\newcommand{\KK}{\mathbf{K}}
\newcommand{\qq}{\mathbf{q}}
\newcommand{\rr}{\mathbf{r}}
\newcommand{\RR}{\mathbf{R}}

\begin{document}

\title{Band Unfolding Made Simple}

\author{Sara G. Mayo, Felix Yndurain and Jose M. Soler}
\affiliation{
 Dpto.\:F\'isica de la Materia Condensada \\
 Universidad Aut\'onoma de Madrid \\
 Cantoblanco, 28049 Madrid, Spain}

\date{\today}

\begin{abstract}
We present a simple view on band unfolding of the energy bands obtained from supercell calculations. It relies on the relationship between the local density of states in reciprocal space (qLDOS) and the \emph{fully unfolded} band structure. This provides an intuitive and valid approach not only for periodic, but also for systems with no translational symmetry. By \emph{refolding} into the primitive Brillouin zone of the pristine crystal we recover the conventional unfolded bands. We implement our algorithm in the  \textsc{Siesta} package and apply it to defects on Si and graphene.
\end{abstract}

\maketitle

%-------------------------------------
\section{\label{sec:intro}Introduction}

Plots of so called energy bands are the most basic and used tool in interpreting the calculated electronic structure of simple crystals.
Such plots represent the energy of the Bloch orbitals as a function of their crystal momentum in the primitive Brillouin zone (PBZ), that is, $E\left(\kk\right)$. 
These theoretical band structures can be obtained within the tight-binding approximation or the density functional theory (DFT) and they have a direct connection with the results of angle resolved photoemission spectroscopy (ARPES) experiments. 
However, the simplicity of this approach disappears when the calculations involve large supercells with many atoms. 
As the size of the cell in real space increases, the first Brillouin zone in reciprocal space shrinks and more lines populate the band structure, hindering the extraction of useful information and the comparison with experiments.

Several authors have already developed techniques to unfold the supercell Brillouin zone into the primitive Brillouin zone. 
Some of the existing works focus on algorithms within the tight-binding approximation \cite{Boykin2005,Boykin2007,Boykin2018,Dargam} or first principles calculations, employing as basis sets linear combinations of atomic orbitals (LCAO) \cite{Lee}, plane waves \cite{Chen2018} or Wannier functions \cite{Ku}. Another methodology studies the electronic structures of alloy systems \cite{Popescu10,Popescu12}. 
Other authors have delved into the theory of the problem and they have developed general formulations of it \cite{Allen2013,Huang,Medeiros,Rubel,Kosugi}.
%Some works focus on studying the bands of alloy systems \cite{Popescu10,Popescu12}, whereas others describe algorithms within the tight-binding approximation 
%\cite{Boykin2005,Boykin2007,Boykin2018,Dargam} 
%or first-principles calculations, employing a certain basis set, as linear combination of atomic orbitals (LCAO) \cite{Lee}, plane waves \cite{Chen2018} or Wannier functions \cite{Ku}. 
%Other authors delve into the theory of the problem and develop more general formalisms  \cite{Allen2013,Huang,Medeiros,Rubel,Kosugi}.

Most of these methods search a direct relationship between Brillouin zones of the primitive crystal cell and of the simulation supercell, by expanding the supercell states in a basis set with the periodicity of the crystal. 
In contrast, we aim to simplify the procedure and provide a link between existing methods by dividing the unfolding problem in two steps. First, we consider the \textit{fully unfolded bands}, extended to the whole reciprocal space, through a Fourier decomposition of the Bloch wave functions of the system \cite{Kosugi}. This results in a non-periodic description, interesting by itself. Second, by what we call a \textit{refolding} into the crystal primitive Brillouin zone, we recover the conventional unfolded bands of other authors.

\section{\label{sec:method}Method}

\begin{figure}[!ht]
\centering
\includegraphics[width=89mm]{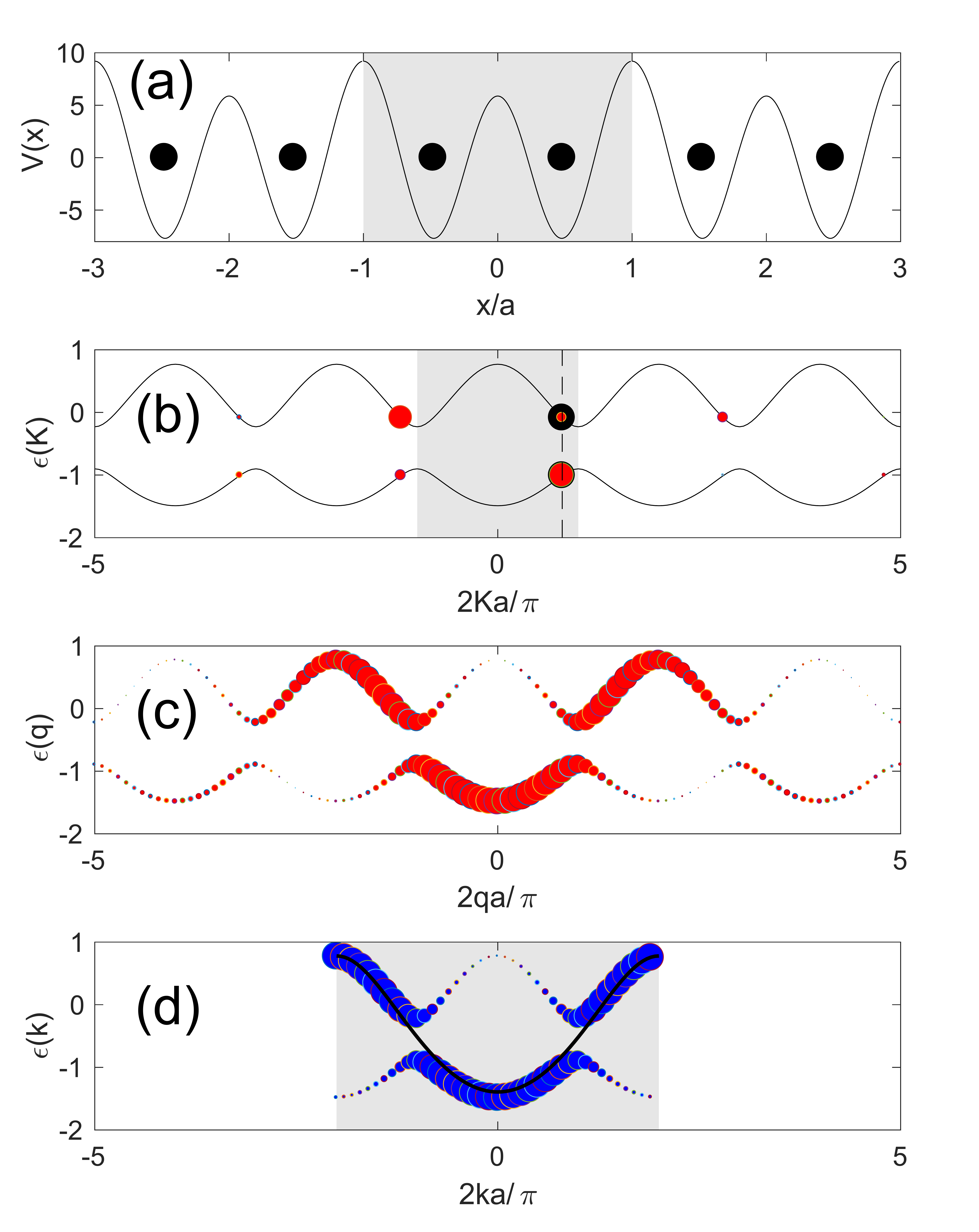}
\caption{\small 
   Scheme of the unfolding method.
   \textbf{(a)} 1D model of a chain of pseudo-atoms (dots) with an attractive gaussian potential (line). The atoms are paired, with a Peierls distortion of 2\% relative to their undistorted distance $a$. The energy origin is the average potential and its units are $\hbar^2 a^2/m_e$. The unit cell of the distorted chain is shaded. 
   \textbf{(b)} First two bands of the distorted chain, with its PBZ shaded. A random $K$ point is singled out (dashed line). The (normalized) weights of the Bloch states at this $K$ (black dots), are split according to the squared Fourier coefficients of their respective wave functions (red dots), appearing at $q=K+G$. %, with $G=\pi n/a$. 
   Dot areas are proportional to weights.
   \textbf{(c)} Fully unfolded bands, obtained through the procedure described in (b) for all $K$s in the PBZ. %These $K$ are continuous, but they have been discretized for clarity. 
   The non-periodic unfolded weights are the local DOS in Fourier space (qLDOS) $n(q,\epsilon)$.
   \textbf{(d)} Bands refolded into the PBZ of the undistorted chain (shaded region). The weight at each $k$ (blue dots) is the sum over $g$ of the unfolded weights in (c), at points $q=k+g$. %, with $g=2\pi n/a$. % reciprocal vectors of the refolding zone. 
   %Since $q$ vectors are continuous, the refolding zone needs not to be a multiple of the zone in (b), although it is so in this example. 
   The continuous line is the first band of the undistorted chain.
}
\label{fig:bandscheme}
\end{figure}

The basic steps of our unfolding method are schematically summarized in Fig.\,\ref{fig:bandscheme}. 
We start by considering the energy bands not just as dispersion relations, but as the density of states in the (first) simulation Brillouin zone (SBZ), the reciprocal of the real-space simulation (super) cell (SC):
% ---  
  \begin{equation}
  n_{SBZ}(\KK,\epsilon) = 
  \sum_i \delta(\epsilon-   \epsilon_{\KK,i})
  \label{nKeps}
  \end{equation}  % \frac{1}{(2\pi)^3}~
% ---
where $\KK$ is a wave vector in the SBZ, and $\epsilon_{\KK,i}$ is the eigenvalue of its $i$th Bloch state, that is, band energy \cite{notation}.

   The normalization of $n_{SBZ}(\KK,\epsilon)$, as well as those of other densities 
   %$n(\qq,\epsilon)$, $n(\epsilon)$, and $n_{RBZ}(\kk,\epsilon)$ 
defined below, is such that they are densities of states {\it per unit of macroscopic volume} (as well as per unit of their respective independent variables), what facilitates the comparison between simulation cells of different volumes.

   Next, we split the normalized weight $\delta(\epsilon-\epsilon_{\KK,i})$ according to the squared Fourier coefficients of the corresponding wave function. Summing over Bloch states, we construct the {\it fully unfolded bands} as the spectral density, which can also be considered as the local density of states in reciprocal space (qLDOS):
% ---
  \begin{eqnarray}
  n(\qq,\epsilon) &=& 
  \sum_i \int_{SBZ} d\KK \: 
  |\tilde{\psi}_{\KK,i}(\qq)|^2 \: 
  \delta(\epsilon-\epsilon_{\KK,i}) \nonumber \\
      &=& \sum_i \: 
      |\tilde{u}_{\KK_\qq,i,\GG_\qq}|^2 \:
       \delta(\epsilon-\epsilon_{\KK_\qq,i}),
  \label{qLDOS}
  \end{eqnarray}  %\frac{1}{(2\pi)^3}~
% ---
where $\psi_{\KK,i}(\rr)$ is a Bloch wave function (normalized in the SC), $u_{\KK,i}(\rr) \equiv \psi_{\KK,i}(\rr) e^{-i\KK\rr}$ is its periodic part, and $\tilde{\psi}_{\KK,i}(\qq), \tilde{u}_{\KK,i,\GG}$ are their respective Fourier transforms:

% ---
\begin{equation}
\psi_{\KK,i}(\rr) = \frac{1}{(2\pi)^{3/2}}\int_\infty d\qq ~\tilde\psi_{\KK,i}(\qq) ~e^{i\qq\rr},
\label{psiR}
\end{equation}
% ---
\begin{equation}
\tilde{\psi}_{\KK,i}(\qq) = (2\pi)^{3/2}~\sum_\GG \:\delta(\KK+\GG-\qq) \:\tilde{u}_{\KK,i,\GG},
\label{psiKjQ}
\end{equation}
% ---
\begin{equation}
\tilde{u}_{\KK,i,\GG} = \frac{1}{V_{SC}} \int_{SC} d\rr ~u_{\KK,i}(\rr) \:e^{-i \GG \rr},
\label{uKiG}
\end{equation}
% ---
with $V_{SC}$ the volume of the SC and $\GG$ its reciprocal wave vectors. 
In Eq.\,(\ref{qLDOS}), $\KK_\qq$ and $\GG_\qq$ are the unique vectors such that: $\KK_\qq$ is within the SBZ;
$\GG_\qq$ is a reciprocal wave vector; and $\KK_\qq+\GG_\qq=\qq$. 
A state $\psi_{\KK,i}$ will contribute to $n(\qq,\epsilon)$ at points $\qq=\KK+\GG$ for all $\GG=\pi N/a$ vectors, due to Bloch's theorem. 
We emphasize that $\qq$ extends to infinity and $n(\qq,\epsilon)$ is not periodic in $\qq$: although the energies at which $n(\qq,\epsilon)$ can be nonzero are periodic, these ``bands" have a different weight at each Brillouin zone (Fig.\,\ref{fig:bandscheme}(c)).

   As can be seen by comparing with Eq.\,(\ref{qLDOS}), the qLDOS is the Fourier-space equivalent of the real-space local density of states (rLDOS), 
\begin{equation}
n(\rr,\epsilon) = \frac{V_{SC}}{(2\pi)^3}~\sum_i \int_{SBZ} d\KK \: |\psi_{\KK,i}(\rr)|^2 \: \delta(\epsilon-\epsilon_{\KK,i}).
\label{rLDOS}
\end{equation}
% ---
The total density of states (DOS) can be obtained by integration of either $n(\rr,\epsilon)$ or $n(\qq,\epsilon)$:
\begin{equation}
n(\epsilon) = \frac{1}{V_{SC}} \int_{SC} d\rr \:n(\rr,\epsilon) =  \frac{1}{(2\pi)^3} \int_\infty d\qq \:n(\qq,\epsilon).
\label{DOS}
\end{equation}

   Since $|\tilde{\psi}_{\KK,i}(\qq)|^2$ is the probability of measuring momentum $\qq$ of a given electron, $n(\qq,\epsilon)$ is the probability of finding an electron (or an empty state) in the system with energy $\epsilon$ and momentum $\qq$, and it can thus be directly related with ARPES results if matrix element effects are taken into account \cite{Kosugi,surface,blackphos}. 
   
The qLDOS, that we call \textit{fully unfolded bands} is the same as the \textit{spectral weight} of other references \cite{Allen2013,Medeiros,Rubel} 
and the \textit{plane-wave unfolded spectra} introduced by Kosugi \textit{et al.} \cite{Kosugi}. Therefore, our approach is a different description, rather than a new method that yields different results. Our emphasis is to provide a clear and simple link with previous methods through the (L)DOS, as well as to generalise band structure analysis to non periodic systems.

   The last step in our method is to {\it refold} $n(\qq,\epsilon)$ into a \textit{refolding Brillouin zone} (RBZ) as
\begin{equation}
n_{RBZ}(\kk,\epsilon) = \sum_\gg n(\kk+\gg,\epsilon),
\label{nRBZ}
\end{equation}
where $\kk$ is within the RBZ and $\gg=\pi n/a$ are its reciprocal lattice vectors. 
Notice that, since $\sum_\GG |\tilde{u}_{\KK,i,\GG}|^2 = 1$, then \: $\sum_\GG n(\KK+\GG,\epsilon) = n_{SBZ}(\KK,\epsilon)$, that is, refolding $n(\qq,\epsilon)$ back into the SBZ recovers the original bands.

Frequently, the simulation cell will be a supercell of the refolding cell. 
   In these cases, the RBZ will be a supercell of the SBZ, and vectors $\gg$ will belong to the set of $\GG$s (Fig.\,\ref{fig:bandscheme}(d)).
   Nevertheless, this condition is not necessary in our method, and in fact it will not be true in many cases, as for simulation cells of liquids or amorphous systems, or of defects that induce strong deformations \cite{cost}.    

The above full unfolding/refolding method can be immediately applied in a plane wave DFT code \cite{Abinit,QE,Vasp}, since the Fourier coefficients of the Bloch wave functions are then directly available. 
The slow decay with momentum of all-electron wavefunctions can be addressed by using pseudopotentials or by introducing a momentum cutoff.
For a basis of atomic orbitals, we expand the Bloch states as
\begin{equation}
\psi_{\KK,i}(\rr) = \sum_\RR \sum_\mu c_{\KK,i,\mu} \:\phi_\mu(\rr-\RR-\rr_\mu)
\: e^{i \KK (\RR+\rr_\mu)},
\label{psiRexpand}
\end{equation}
where $c_{\KK,i,\mu}$ are expansion coefficients and $\phi_\mu$ are atomic orbitals centered at position $\RR+\rr_\mu$ ($\RR$ being SC lattice vectors).
   Substituting into Eq.\,(\ref{uKiG}) we find
\begin{equation}
\tilde{u}_{\KK,i,\GG} = \frac{(2 \pi)^{3/2}}{V_{SC}} \sum_\mu c_{\KK,i,\mu} 
\:\tilde{\phi}_\mu(\KK+\GG) ~e^{-i \GG \rr_\mu},
\label{uG}
\end{equation}
where $\tilde{\phi}_\mu(\qq)$ is the Fourier transform of the numerical atomic orbital $\phi_\mu(\rr)$, with well defined angular momentum quantum numbers $(l_\mu,m_\mu)$, that can be decomposed into radial and angular parts:
\begin{equation}
\phi_\mu(\rr) = \phi_\mu(r) \:Y_{l_\mu,m_\mu}(\hat{\rr})
\label{phiR}
\end{equation}
\begin{equation}
\tilde{\phi}_\mu(\qq) = \tilde{\phi}_\mu(q) \:Y_{l_\mu,m_\mu}(\hat{\qq})
\label{phiFT}
\end{equation}
\begin{equation}
\tilde{\phi}_\mu(q) = \sqrt{\frac{2}{\pi}} (-1)^{l_\mu} \int_0^\infty r^2 dr \:j_{l_\mu}(qr) \:\phi_\mu(r),
\label{phiRadFT}
\end{equation}
%---
with $Y_{l,m}(\hat{\rr})$ spherical harmonics and $j_l(x)$ spherical Bessel functions. 

This algorithm has been implemented in the \textsc{Siesta} package  \cite{Soler2002}.
After a converged DFT \textsc{Siesta} calculation, the hamiltonian and overlap matrices, in the atomic basis set, are calculated and written in a file.
This file, as well as those specifying the radial numerical atomic orbitals, are read by an external utility program that calculates the fully unfolded and refolded spectra at the desired $\qq$ and $\kk$ band lines.
Some \textsc{Siesta} subroutines are also used by the unfolding/refolding program to obtain the wave function coefficients at each required $\KK$ point of the SBZ, as well as to perform the Fourier transforms in Eq.\,(\ref{phiRadFT}).

\section{\label{sec:app}Applications}

We apply our previously described \textsc{Siesta} implementation to a Si FCC crystal with a single vacancy, a model of amorphous Si (a-Si), a monolayer of graphene with a (585) divacancy, and a rotated graphene bilayer under pressure.
We employed the GGA-PBE \cite{PBE} functional for exchange and correlation and double-$\zeta$ + polarization (DZP) basis sets (double-$\zeta$ (DZ) for a-Si).

\subsection{\label{sec:def}Vacancy in Si crystal}

\begin{figure}[h!]
\centering
\subfigure{\includegraphics[width=91mm]{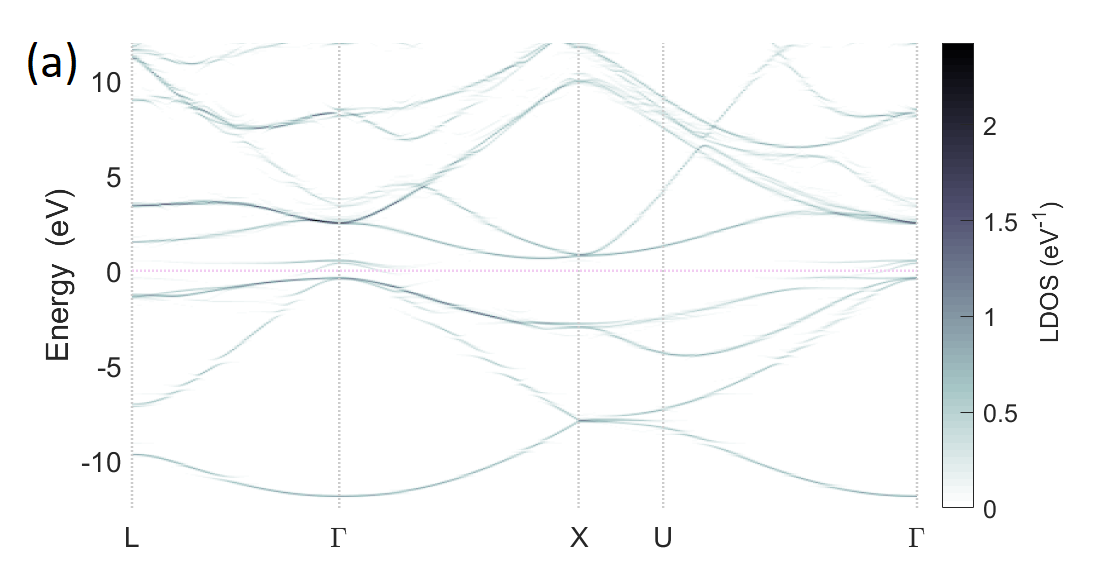}}
\subfigure{\includegraphics[width=86mm]{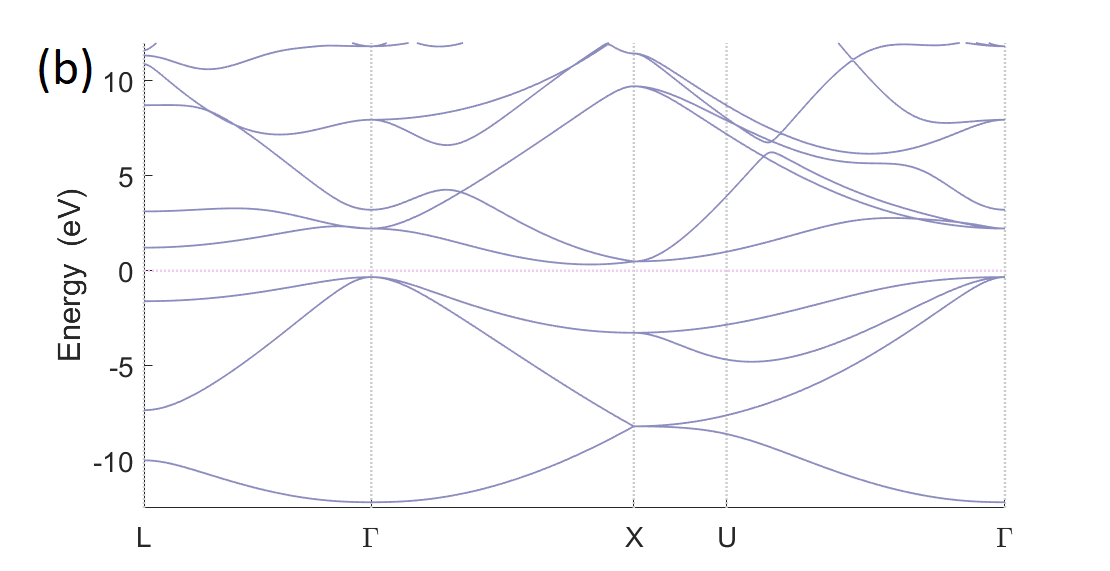}}
\subfigure{\includegraphics[width=88mm]{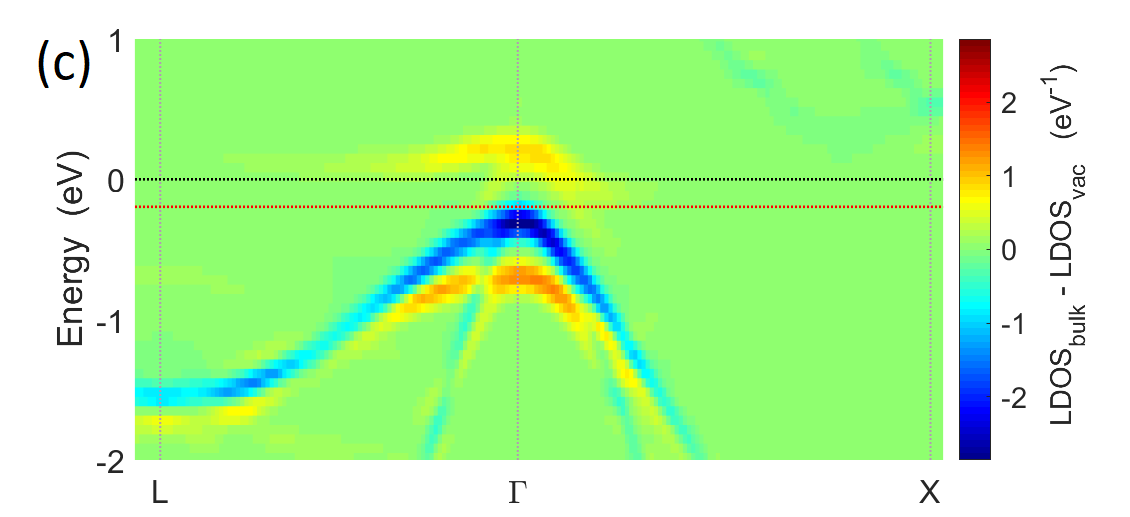}}
\caption{\textbf{(a)} Refolded bands, into the crystal PBZ, of a single vacancy in bulk Si. Energies are relative to the Fermi level, set at zero (horizontal dotted line). Vertical dotted lines mark high symmetry points of the PBZ. The colorbar is the scale for the LDOS.; 
\textbf{(b)} (conventional) bands of Si FCC crystal; 
\textbf{(c)} difference between the refolded LDOS of the defective Si and the pristine crystal, zoomed in the gap region, near the $\Gamma$ point. States arising (vanishing) due to the vacancy appear in hot (cold) colours. The black (red) dashed line is the Fermi level of the perfect (defective) system. 
\small}
\label{fig:vac}
\end{figure}

We model a vacancy in a Si FCC crystal using a 63-atom supercell. 
Its refolded bands into the crystal PBZ vectors are depicted in Fig.\,\ref{fig:vac}(a), and can be compared with the bands of the periodic crystal, in Fig.\,\ref{fig:vac}(b). Changes are appreciated at a careful sight. The refolded bands become blurred and widened due to the appearance of small splittings. Some of these are a consequence of the supercell approach and they become smoother in a larger SC. 

The most relevant changes occur around the Fermi level. A new state arises within the gap, with higher weight around the $\Gamma$ point. The top of the valence band, around $\Gamma$, decreases in energy and in weight. 
We show the difference between the crystal and defective refolded spectra in Fig.\,\ref{fig:vac}(d) around $\Gamma$, at the gap, to remark these changes.

\subsection{\label{sec:am}Amorphous Si}

A clear example of disorder is an amorphous solid. In this case, we cannot talk about a proper band structure, but yet the energy dispersion of the electron states provides interesting results. 
We studied the fully unfolded and refolded bands of a-Si using supercells of 216, 512 and 1024 atoms, modelled by Igram \emph{et al.} \cite{Igram}, obtaining equivalent results for all cases. We present the results for the 512-atom supercell. 

\begin{figure}[h!]
\centering
\subfigure{\includegraphics[width=89mm]{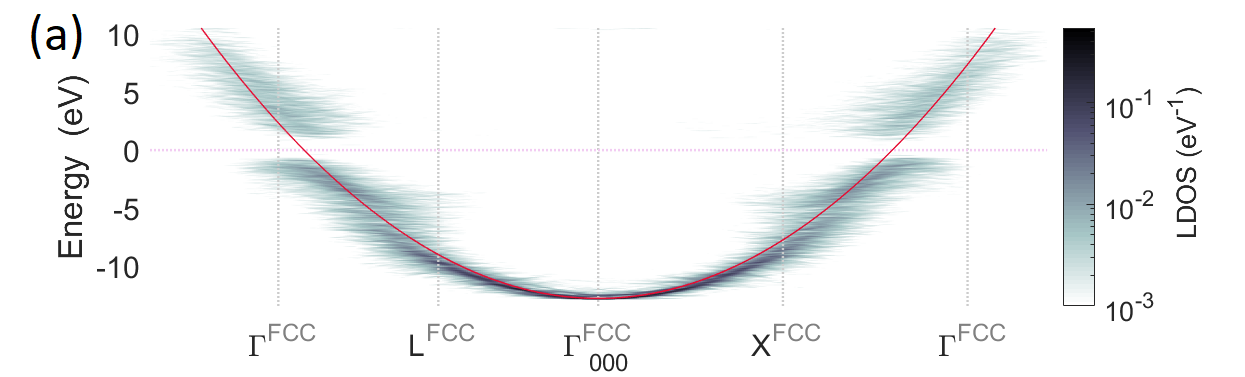}}
%\subfigure{\includegraphics[width=89mm]{05gammaX.png}}
\subfigure{\includegraphics[width=89mm]{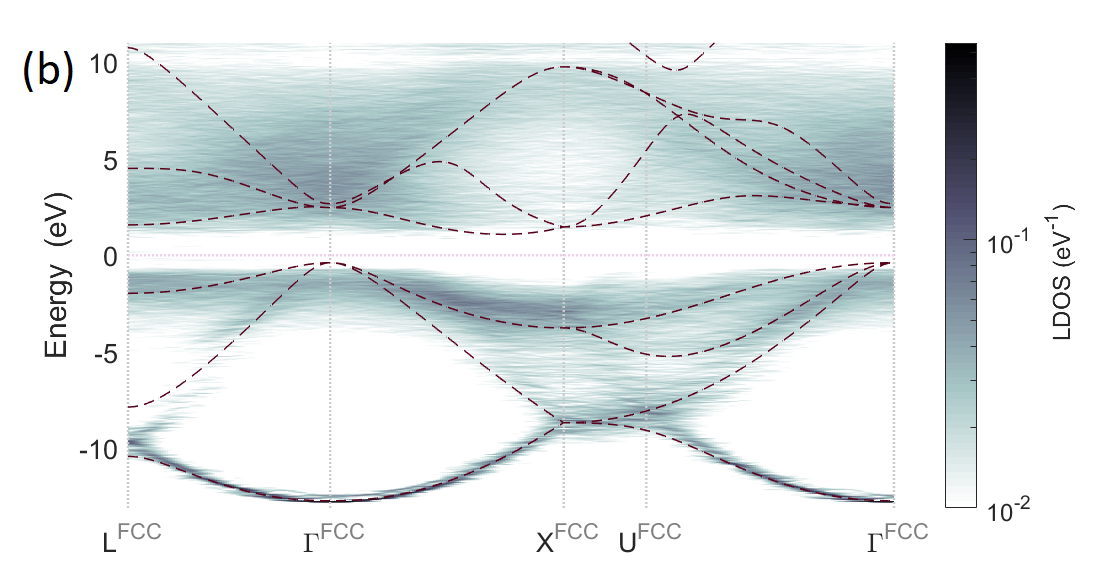}}
\caption{ \textbf{(a)} Fully unfolded bands of a 512-atom cubic cell of amorphous silicon, along the (111) and (100) directions. Vertical lines indicate high symmetry points of the first and second Brillouin zones of the crystal. They are shown for reference but they are not special in the a-Si simulation cell. The red line is a fit to a free-electron dispersion relation with an effective mass $m^*=1.1$; \textbf{(b)} refolding of the a-Si bands into the PBZ of crystalline silicon. Dashed lines are the crystal bands.
\small}
\label{fig:a512}
\end{figure}

Fig.\,\ref{fig:a512}(a) show the fully unfolded bands along two directions of reciprocal space, corresponding to the $L-\Gamma$ and $\Gamma-X$ directions of the crystal, up to the second Brillouin zone. 
As expected from the isotropic character of the amorphous solid, its unfolded bands are essentially identical in both directions, yielding a widened free-electron dispersion with an effective mass $m^*=1.1$ and a gap of $1.5$ eV, consistent with existing values \cite{Igram,Barber}.
%The states with lowest energy have significant Fourier components at low momenta $\qq$, both magnitudes increasing with a free electron dispersion relation behaviour, $\sim\kk^{2}$, modified by the atomic potential.
%We fitted these fully unfolded spectra to a free electron dispersion relation, with a value for the effective mass of $m^*=1.1$ in both directions.

The same bands, refolded to the PBZ of the crystal, are shown in Fig.\ref{fig:a512}(b). At close magnification and inspection, a few localised states appear in the gap due to defects in the a-Si model \cite{Igram}.  % .
% https://www.sciencedirect.com/science/article/pii/0038110167901220
Interestingly enough, despite its isotropic and non-periodic structure, and the incommensurability of its simulation cell with that of the crystal, the refolded bands of a-Si appear as a blurred version of the crystalline silicon (c-Si) bands, specially at low energies. Also, higher LDOS of a-Si can be appreciated in regions of c-Si band degeneracies. This similarity is independent of the size of the a-Si simulation cell employed. We attribute it to the similarity of the local structure in a-Si and c-Si %, with very similar 
in bond distances and angles \cite{Igram}.

\subsection{\label{sec:graphene}Divacancy in graphene}

Graphene is a material with unique electronic properties, but highly sensitive to structural disorder. The presence of defects leads to significant changes on its bands, specially around the Fermi level. Many types of defects have been studied by theorists, such as adatoms, vacancies or Stone-Wales defects, with the aim of predicting their properties and, potentially, using them to tailor the functionalities of graphene. 
Hence, unfolding their band stuctures will shed light on how they modify the original graphene bands.

\begin{figure}[h!]
\centering
\subfigure{\includegraphics[width=43mm]{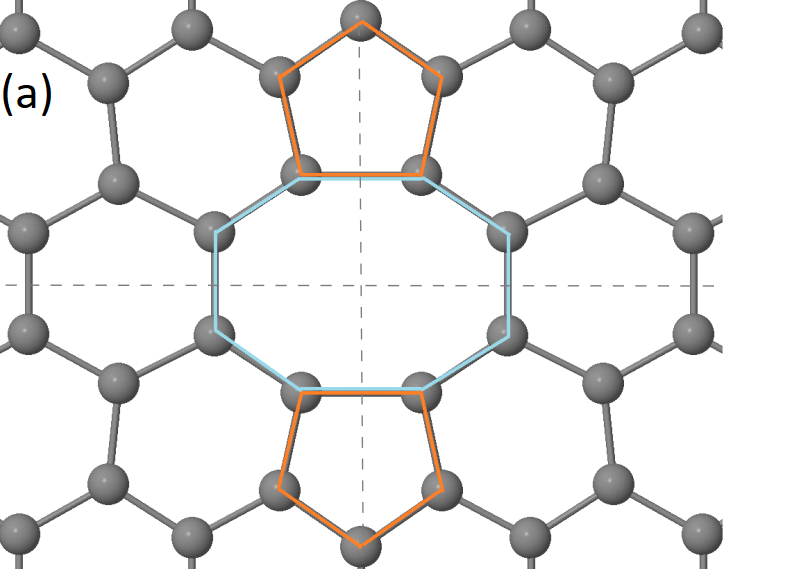}}
\subfigure{\includegraphics[width=35mm]{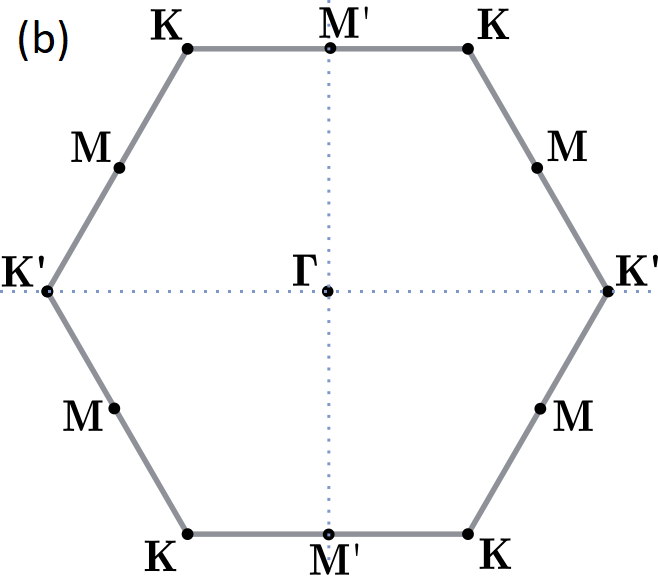}}
\caption{ \textbf{(a)} Relaxed structure of the (585) divacancy in graphene. The dashed lines are the mirror planes of the defect. \textbf{(b)} PBZ of graphene, its symmetry modified by the (585) defect. 
%The dotted lines are the mirror planes of the Brillouin Zone.
\small}
\label{fig:585}
\end{figure}

\begin{figure*}[t]
\centering
\subfigure{\includegraphics[width=59.1mm]{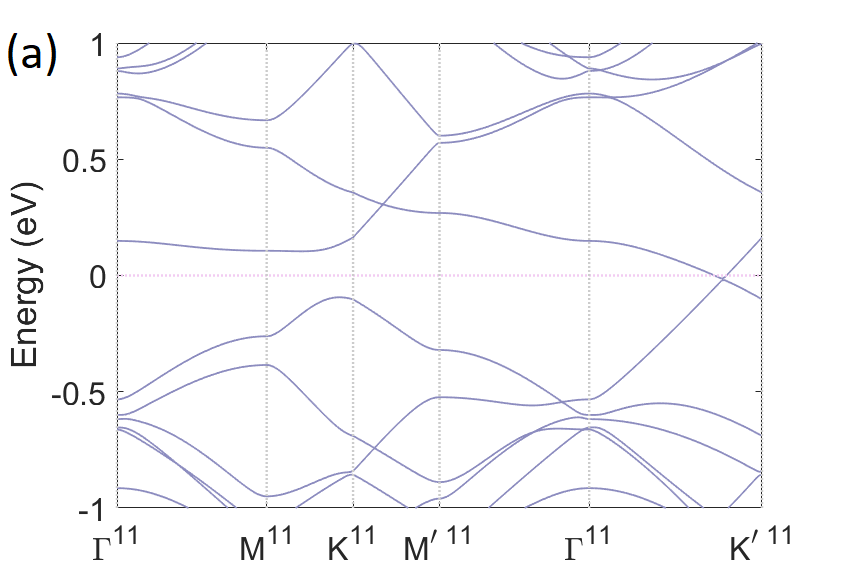}}
\subfigure{\includegraphics[width=59.1mm]{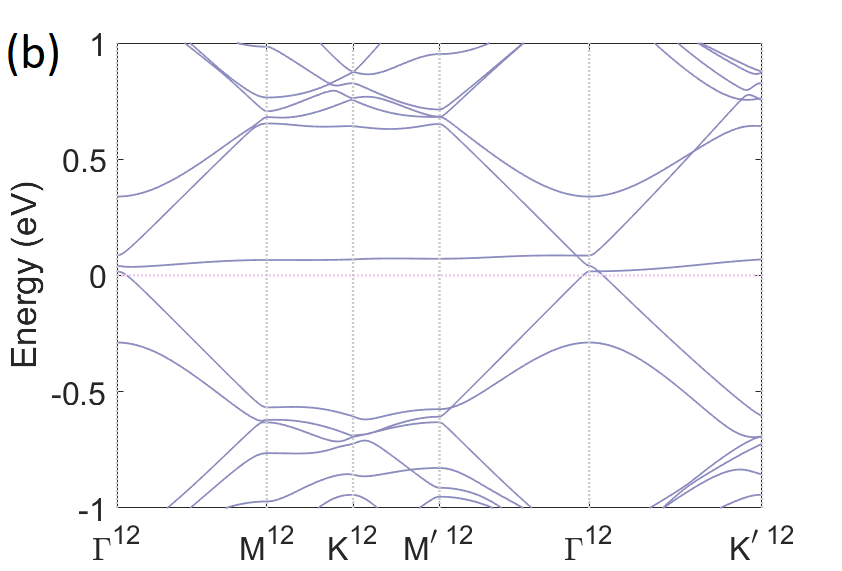}}
\subfigure{\includegraphics[width=59.1mm]{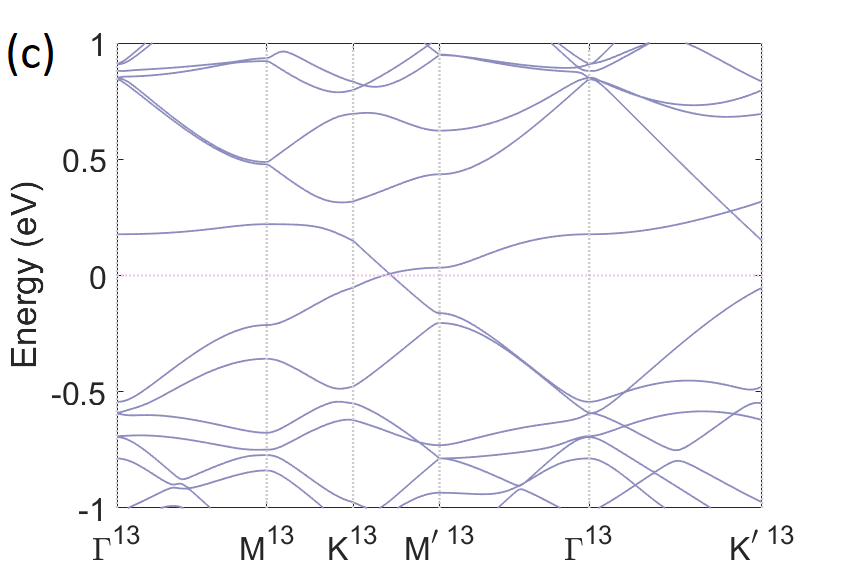}}
\caption{ Conventional bands of a (585) divacancy in graphene modelled in \textbf{(a)} $11 \times 11$, \textbf{(b)} $12 \times 12$ and \textbf{(c)} $13 \times 13$ simulation cells, in their respective SBZs.
\small}
\label{fig:Dbands}
\end{figure*}

\begin{figure*}[t!]
\centering
\subfigure{\includegraphics[width=59mm]{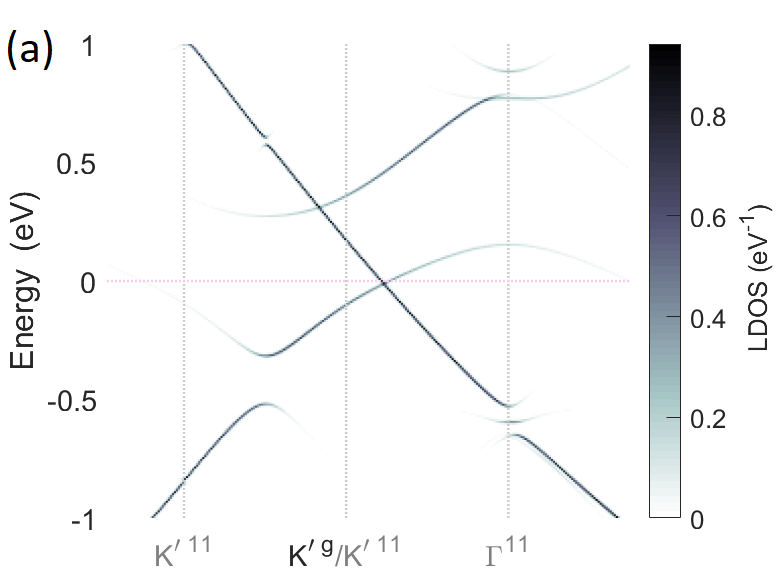}}
\subfigure{\includegraphics[width=59mm]{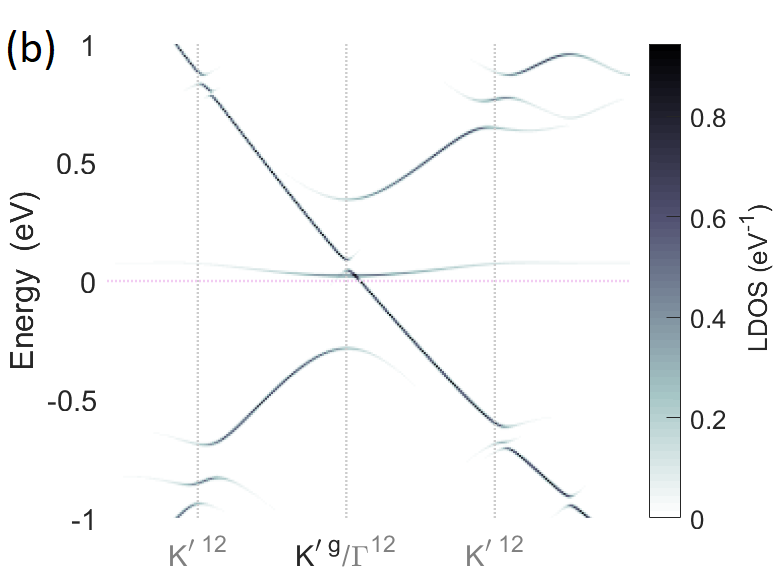}}
\subfigure{\includegraphics[width=59mm]{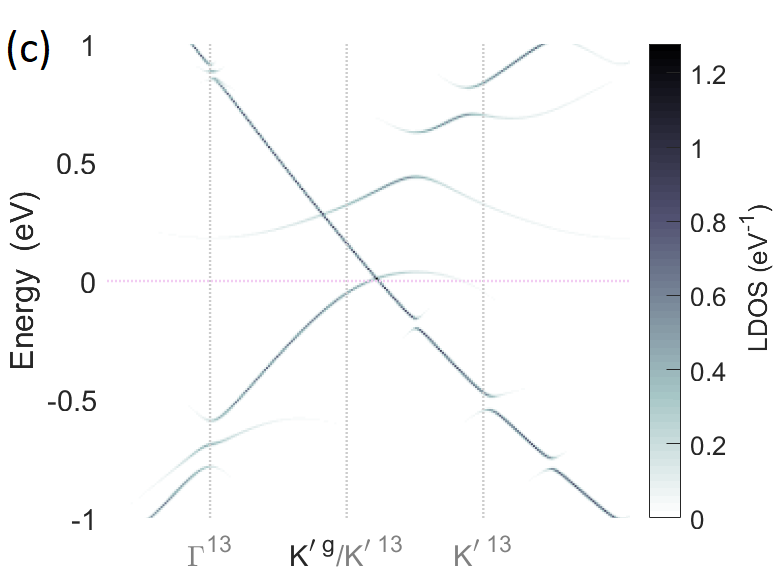}}
%\subfigure{\includegraphics[width=59mm]{11-3D.png}}
%\subfigure{\includegraphics[width=59mm]{12-3D.png}}
%\subfigure{\includegraphics[width=59mm]{13-3D.png}}
\caption{ Bands around the Fermi level of a (585) divacancy in graphene, refolded into the PBZ, obtained from \textbf{(a)} $11 \times 11$, \textbf{(b)} $12 \times 12$ and \textbf{(c)} $13 \times 13$ simulation cells. $K'^g$ denotes the $K'$ point of the PBZ of graphene. $K'^n$ indicates a $K'$ point of the $(n \times n)$ SBZ.
\small}
\label{fig:Dref}
\end{figure*}

\begin{figure*}[htbp!]
\centering
%\subfigure{\includegraphics[width=59mm]{k11.png}}
%\subfigure{\includegraphics[width=59mm]{k12.png}}
%\subfigure{\includegraphics[width=59mm]{k13.png}}
\subfigure{\includegraphics[width=59mm]{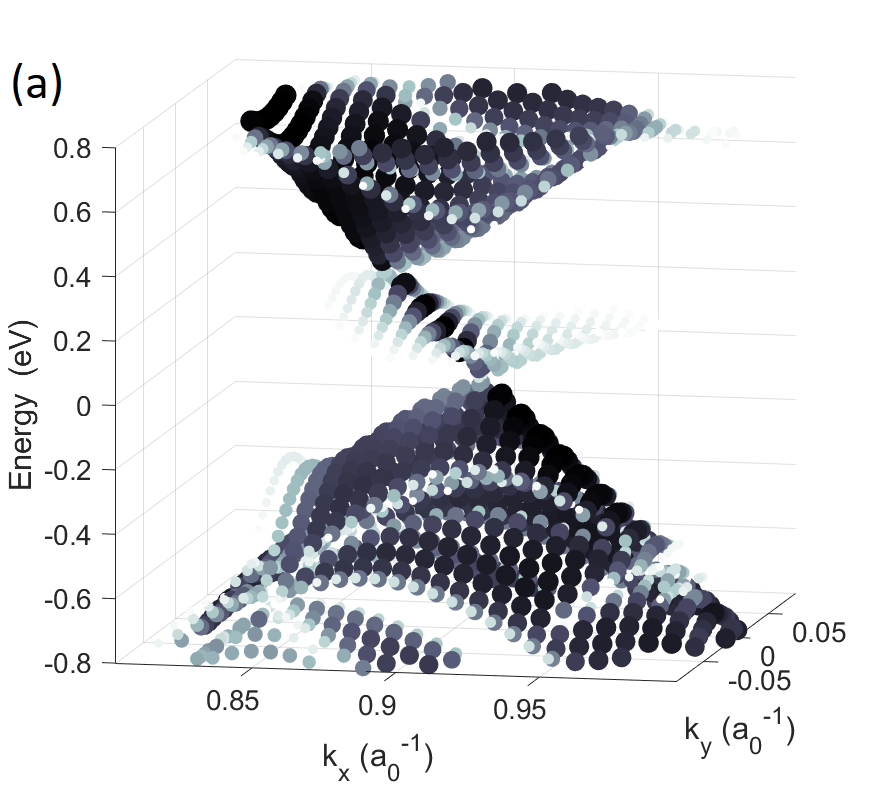}}
\subfigure{\includegraphics[width=59mm]{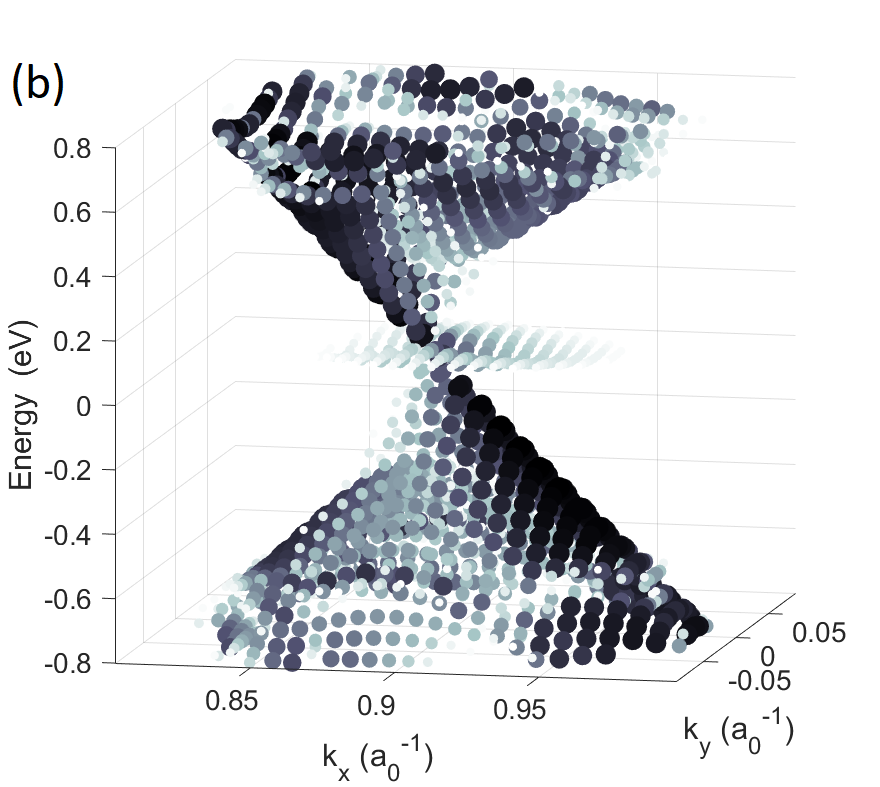}}
\subfigure{\includegraphics[width=59mm]{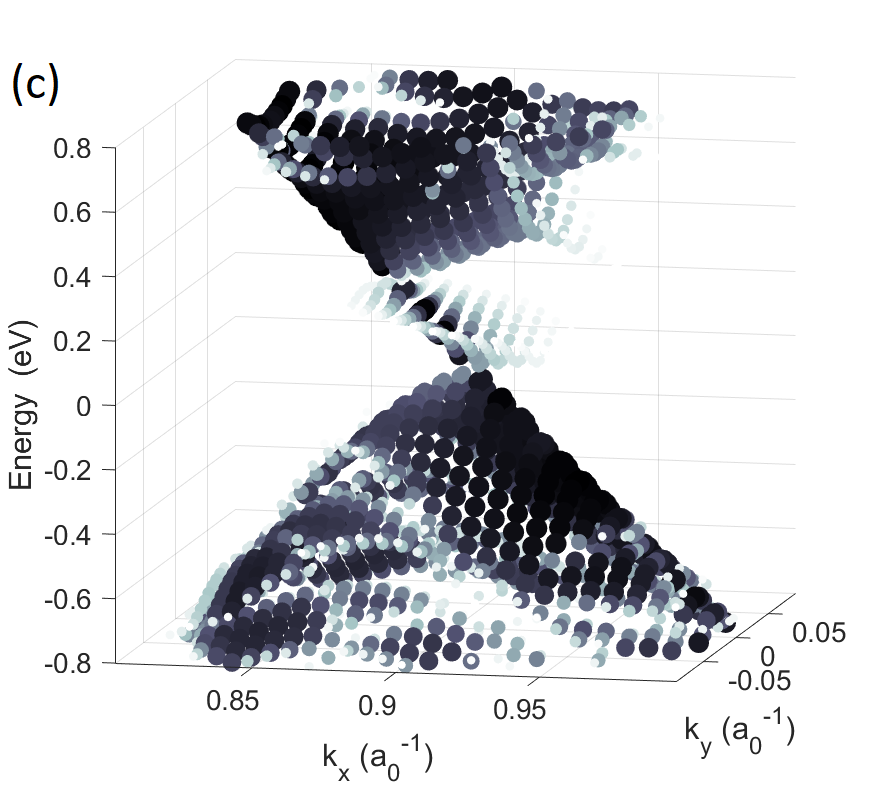}}
\caption{ 2D bands of a (585) divacancy refolded into the PBZ, in the surroundings of the K$'$ point, obtained from \textbf{(a)} $11 \times 11$, \textbf{(b)} $12 \times 12$ and \textbf{(c)} $13 \times 13$ simulation cells. The darkness times size of the dots is proportional to the LDOS. The units for $k$ are $1/a_0$, where $a_0$ is the lattice parameter. (Interactive figures available as supplementary material).
\small}
\label{fig:Dmesh}
\end{figure*}

We consider a (585) divacancy in a graphene monolayer, a defect that has been synthesised and characterised by Ugeda \emph{et al}. \cite{Ugeda}. The (585) defect is formed by two adjacent C vacancies rearranged into two pentagons and one octagon, as shown in Fig.\,\ref{fig:585}(a), with no dangling $\sigma$ bonds.  
We modelled the divacancy in simulation cells of different sizes, from $7 \times 7$ to $14 \times 14$. 
%The obtained results can be divided in two groups: $3n\times3n$ graphene PBZs and other sizes. 
Here we report the $11 \times 11$, $12 \times 12$ and $13 \times 13$ as representative.

The conventional bands of the supercell calculations within their SBZs, Fig.\,\ref{fig:Dbands}, reveal that the former six-fold symmetry of the Brillouin zone is broken, leading to inequivalent $K$ and $M$ points and two emerging mirror planes, Fig.\,\ref{fig:585}(b), as happens in the atomic structure. The $K$ and $K'$ points no longer present a Dirac cone, although all of them present a band crossing of $E_F$ at different points in their Brillouin zones: near $K'$ for the $11 \times 11$ SC, around $\Gamma$ for the $12 \times 12$ SC and between $M'$ and $K$ for the $13 \times 13$ SC. Besides this, it is hard to find similarities between them. 
To relate these overcrowded spectra with the band structure of graphene, we compute the refolded bands into the PBZ in the surroundings of $K$ and $K'$ points of graphene. 

Fig.\,\ref{fig:Dref}(a) and (c) show the $11 \times 11$ and $13 \times 13$ SCs bands refolded into the PBZ.  The Fermi level crossing is located around $K'$ of graphene, which corresponds to a $K'$ point in both SBZs. One of the bands conforming the former Dirac cone stays almost unaltered, while the other one is split, forming two cone tips shifted in the $k_x$ direction and connected by a state. In the 2D bands plot of Fig.\,\ref{fig:Dmesh}(a) and (c) this is seen with more clarity. The Fermi level decreases from its original value, being coincident with the tip of the lower cone.

In the case of the $12 \times 12$ SC, Fig.\,\ref{fig:Dref}(b), we see a similar behaviour, with slight differences. One band of the cone remains almost unaltered, with a small gap of $0.04$ eV opening at $K'$. The other one is split, with one of its fragments conforming a flat state at the Fermi level, leading to a single band crossing. In Fig.\,\ref{fig:Dmesh}(b) this flat state and the cone tips can be appreciated. 
Here, the $K'$ of graphene coincides with a $\Gamma$ point of the $12 \times 12$ SBZ.

Despite the differences between the three sizes, a general trend is clearly identified after refolding into the PZB of graphene, which is not the case in the traditional bands description of Fig.\,\ref{fig:Dbands}. All cases present a splitting and a shift in $k_x$ of the Dirac cone, as well as a single band between the two cone tips as the lowest unoccupied electronic state. The Fermi level is located at the tip of the lower cone. We remark that, in $3n\times3n$ SCs, the $K'$ of the PBZ corresponds to a $\Gamma$ point in the SBZ, unlike in the $(3n+1)\times(3n+1)$ and $(3n+2)\times(3n+2)$ cases. This is consistent with the formal differences existing between both kinds of supercells \cite{Lastra}. 
Finally, the refolded bands obtained around $K$ are inverted in $k_y$ compared to the refolded bands around $K'$ shown above. 

\subsection{\label{sec:graphene}Pressure in rotated graphene bilayers}

Another path to modify the electronic behaviour of graphene is given by rotated graphene bilayers. These are defect-free systems with a Moir\'e pattern, involving large periodicities. At small angles, the interaction between the two layers induces two saddle points in the band structure, along with two logarithmic van Hove singularities in the DOS. As the angle decreases, the singularities approach and, eventually, at the so called magic angles ($\theta=1.1^\circ$), they collapse into a single peak at the Fermi level \cite{Cao}. The same effect has been recently reported to occur for larger angles, when external pressure is applied \cite{Felix,Bezanilla,Carr,Yankowitz}. 

\begin{figure}[b]
\centering
\includegraphics[width=89mm]{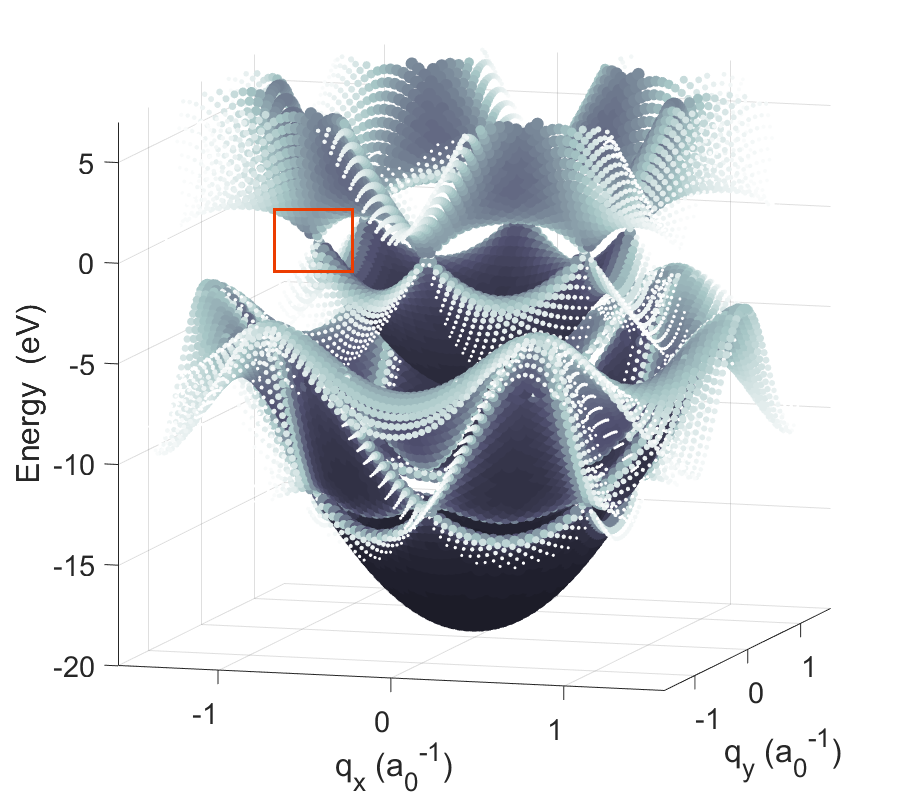}
\caption{
Fully unfolded bands of a graphene monolayer, up to the second Brillouin zone. The squared area marks a Dirac cone, region considered for further study in the bilayer case.
\small }
\label{fig:graphene}
\end{figure}

\begin{figure}[b]
\centering
\subfigure{\includegraphics[width=89mm]{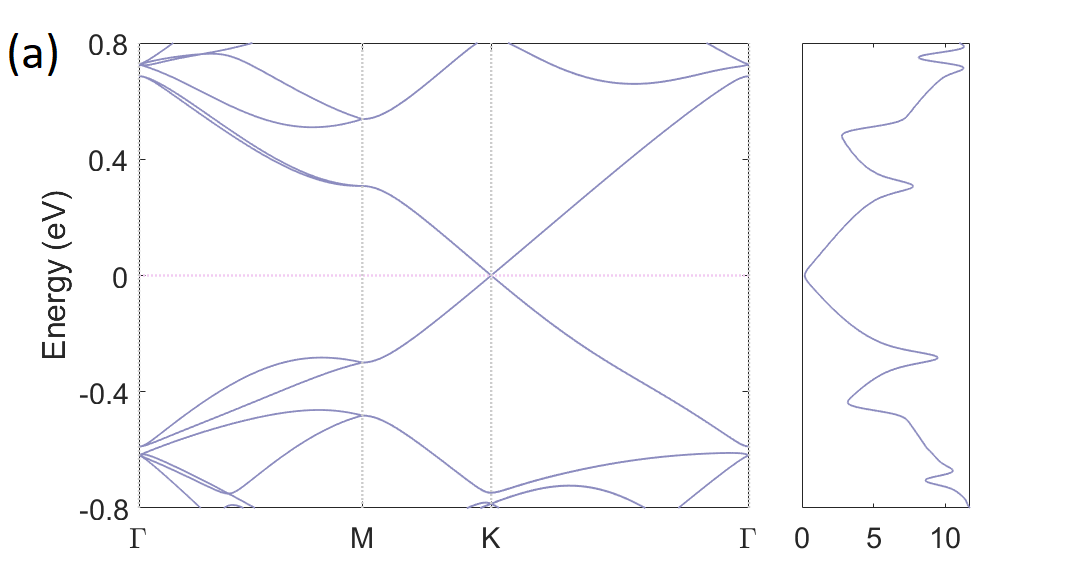}}
\subfigure{\includegraphics[width=89mm]{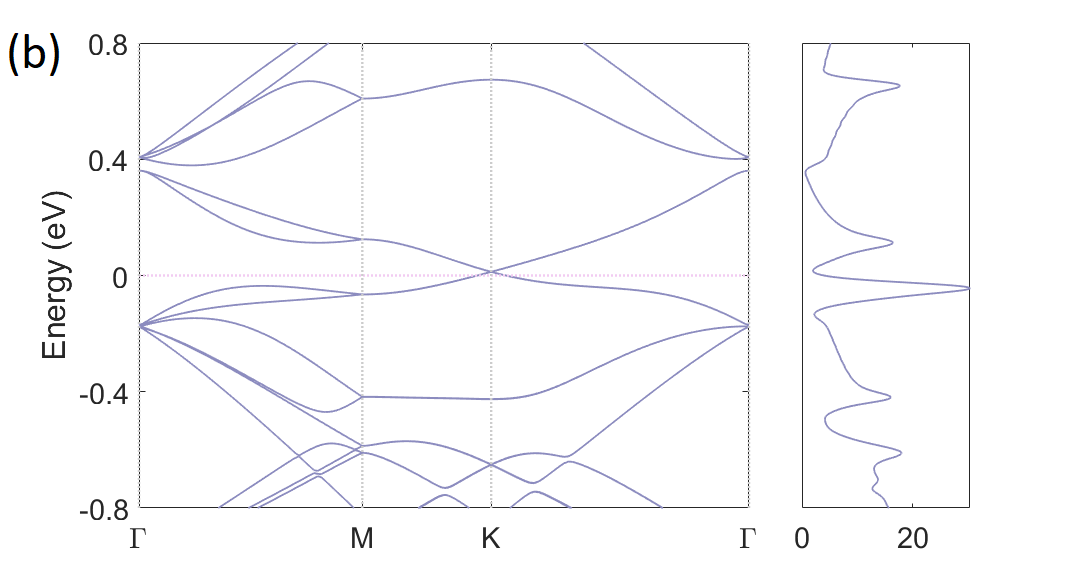}}
\subfigure{\includegraphics[width=89mm]{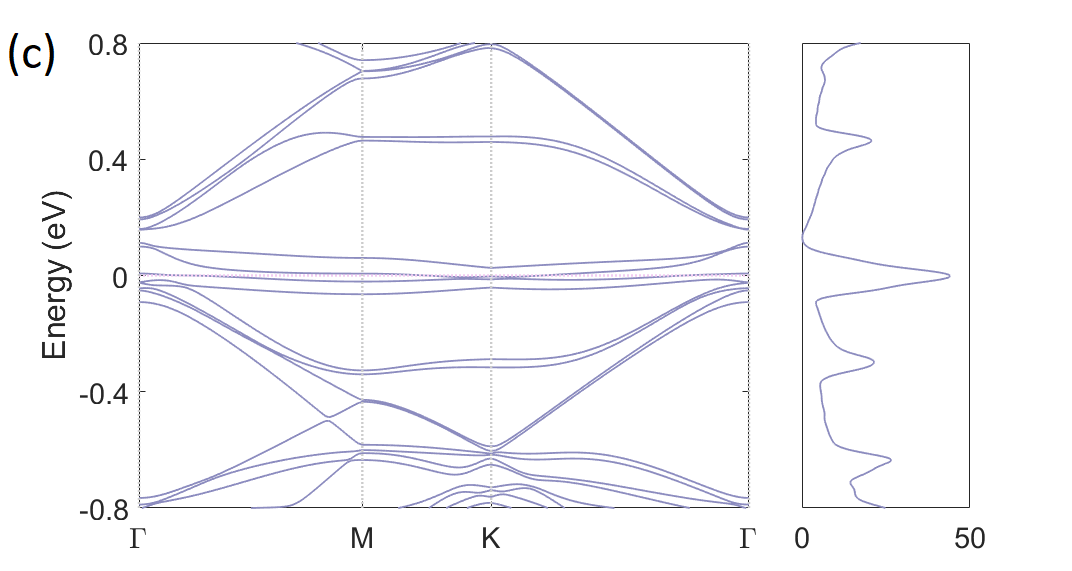}}
\caption{ Conventional bands in the SBZ and DOS of a graphene bilayer rotated an angle of $5.08^{\circ}$, \textbf{(a)} at equilibrium geometry; \textbf{(b)} under a pressure of 0.70 GPa and \textbf{(c)} under 1.63 GPa.
\small}
\label{fig:Bbandsdos}
\end{figure}

\begin{figure*}[htbp!]
\centering
\subfigure{\includegraphics[width=59mm]{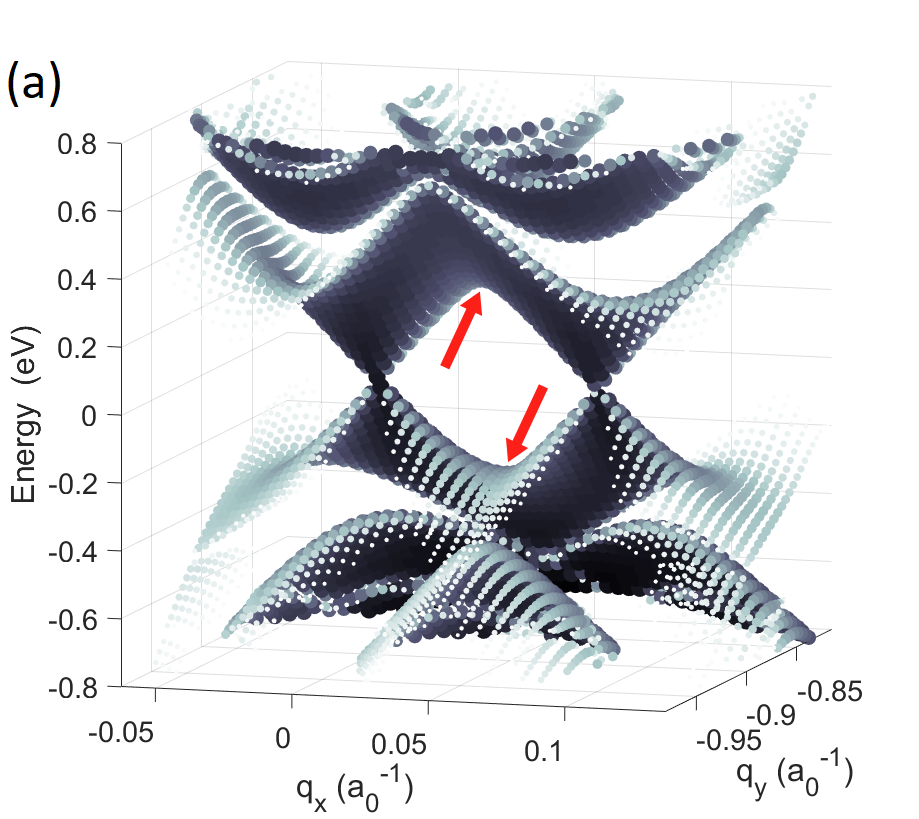}}
\subfigure{\includegraphics[width=59mm]{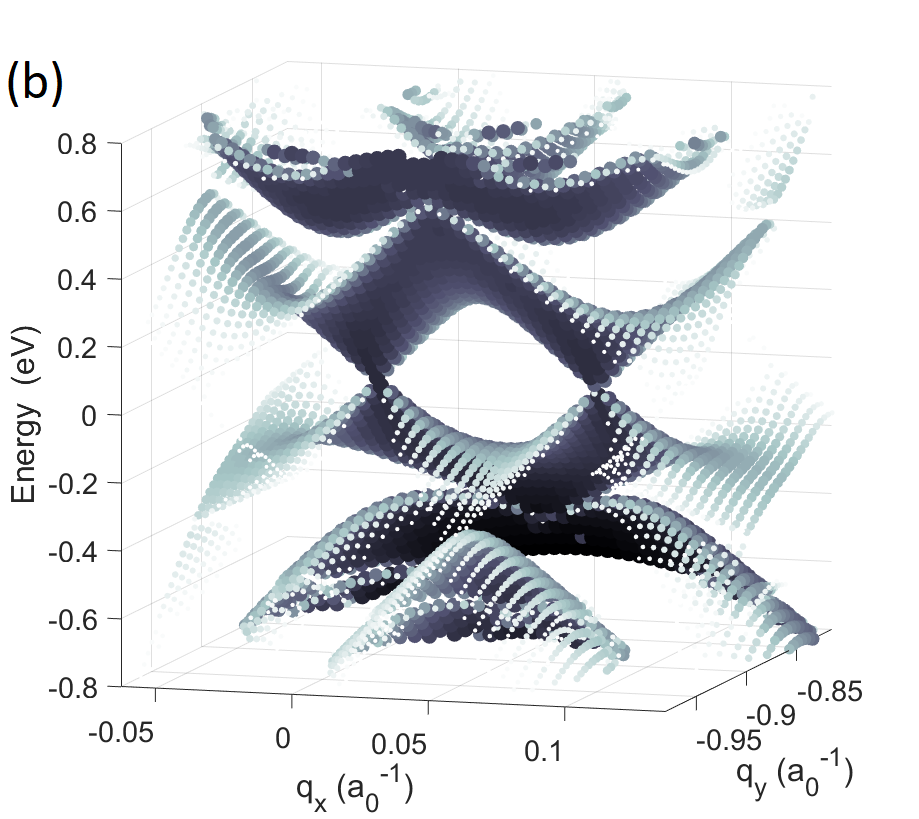}}
\subfigure{\includegraphics[width=59mm]{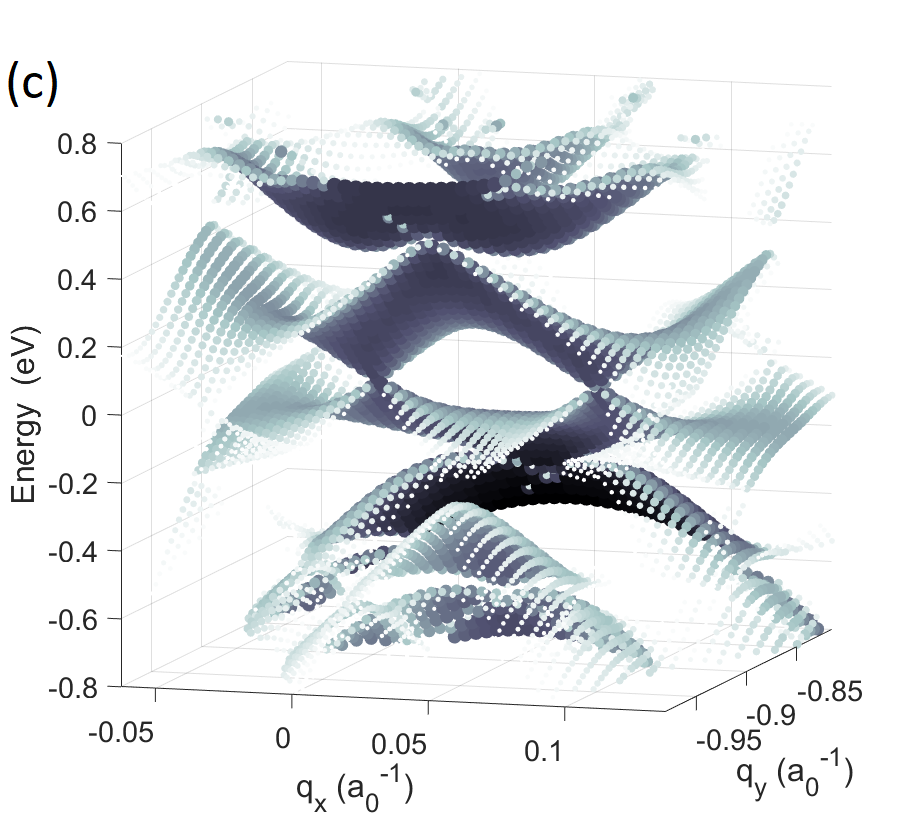}}
\subfigure{\includegraphics[width=59mm]{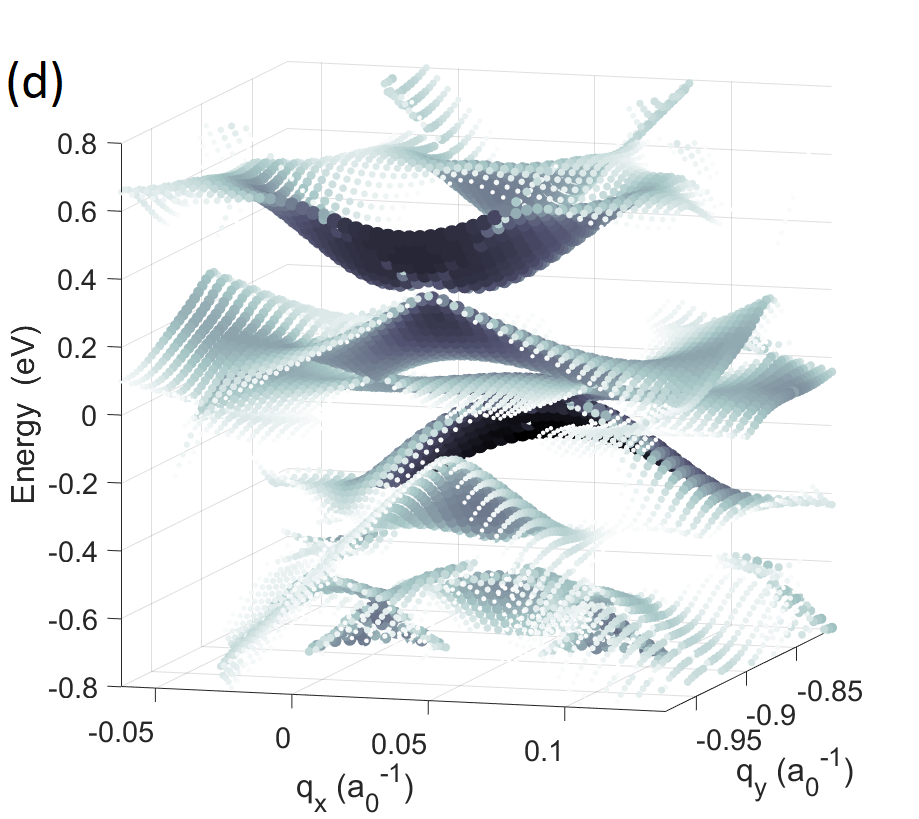}}
\subfigure{\includegraphics[width=59mm]{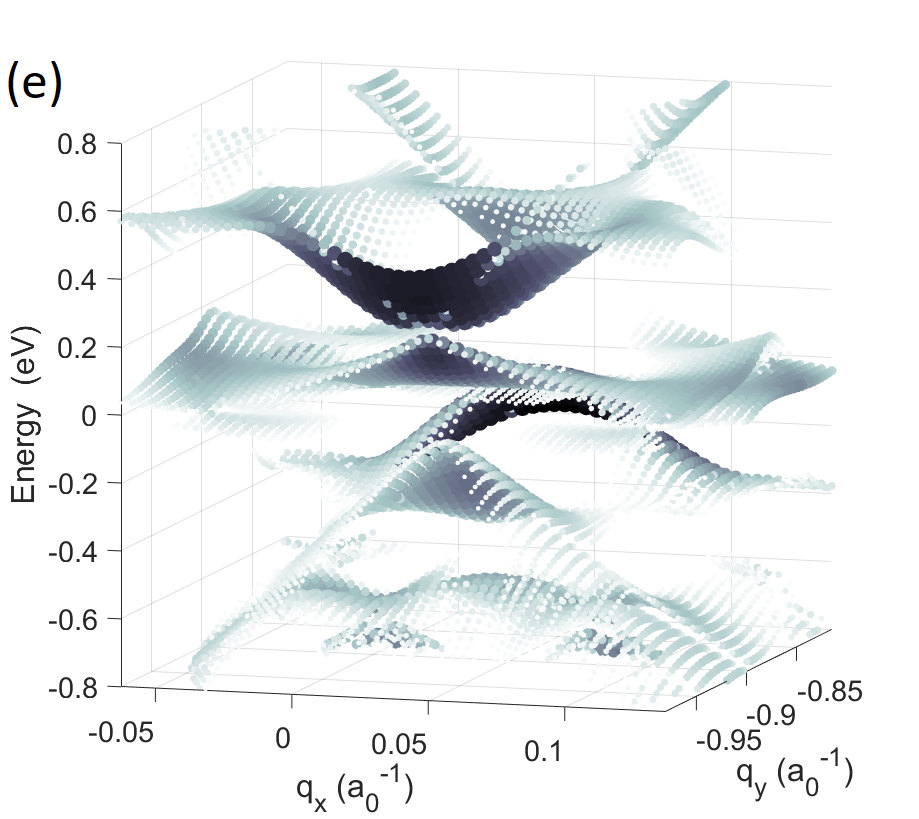}}
\subfigure{\includegraphics[width=59mm]{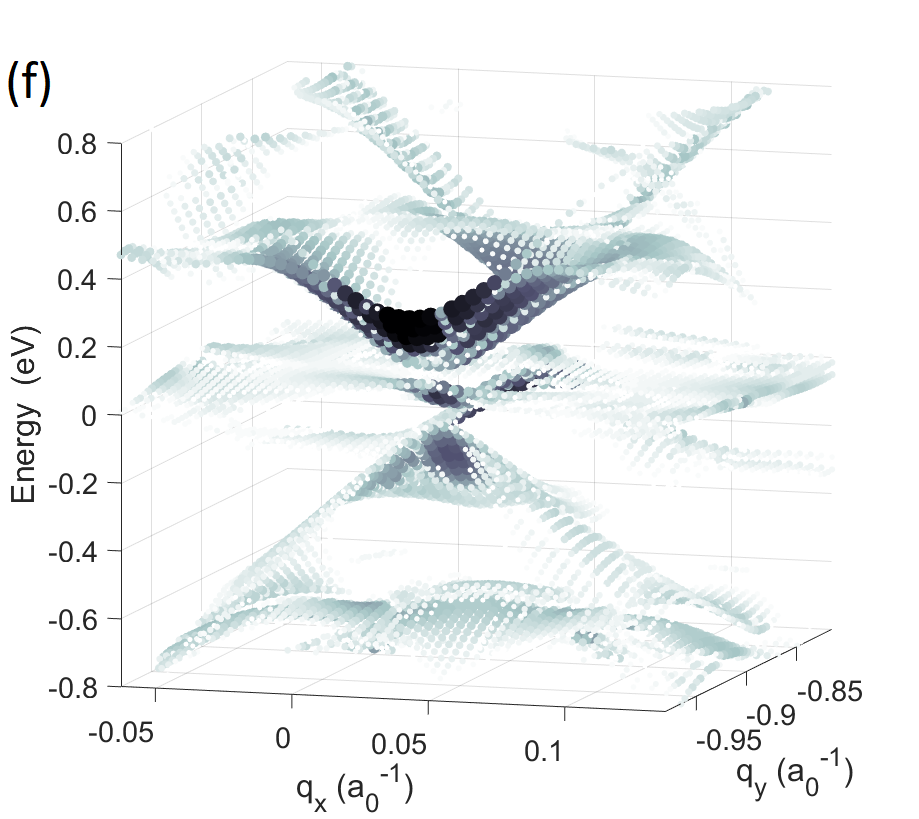}}
\caption{ 2D fully unfolded bands of a graphene bilayer rotated $5.08^{\circ}$, in the surroundings of the first $K$ points of the monolayers, at different pressures: 
\textbf{(a)} equilibrium distance. The arrows indicate the saddle points that originate the van Hove singularities; \textbf{(b)} 0.04 GPa; \textbf{(c)} 0.23 GPa; \textbf{(d)} 0.70 GPa; \textbf{(e)} 1.06 GPa and \textbf{(f)} 1.63 GPa. 
The darkness and size of the dots is the scale for LDOS, relative to the maximum value of each data set. Weights smaller than 100 times the maximum are neglected (interactive figures available as supplementary material).
\small}
\label{fig:Bmesh}
\end{figure*}

We study a bilayer rotated $5.08^{\circ}$, at equilibrium under increasing pressures up to $1.63$ GPa. We employ a GGA exchange-correlation functional including Van der Waals interactions \cite{Berland,Roman}. As none of the monolayers lattice orientations has a prevalence over the other, refolding into the PBZ of one of them is not particularly informative. Therefore, we consider the fully unfolded bands as the adequate tool to analyze this system. 
We first show the fully unfolded bands of a graphene monolayer, Fig.\,\ref{fig:graphene}, up its second Brillouin zone, for comparison purposes. Two paraboloids with gaps opening along them and a six-fold symmetry and are clearly distinguished, and conform the dispersion relations of the $\sigma$ and $\pi$ orbitals. In the case of the bilayer, we will restrict the unfolding region to the surroundings of a $K$ (and $K^{5.08^{\circ}}$) point (red square).

In Fig.\,\ref{fig:Bbandsdos} we show the conventional band structures of the bilayer at the equilibrium distance, at a middle stage and under a pressure of $1.63$ GPa, next to their corresponding DOS. The saddle points and van Hove singularities can be appreciated.

Fig.\,\ref{fig:Bmesh} depicts the evolution of the fully unfolded spectra under increasing pressures. At equilibrium configuration, Fig.\,\ref{fig:Bmesh}(a) shows a neat picture of the interaction between the cones of both monolayers, as well as the saddle point emerging in between. The relative maximum intensities of the LDOS are homogeneous in energies, and the states present high dispersion in energies. As pressure is applied, Fig.\,\ref{fig:Bmesh}(b),(c),(d) and (e), the cones flatten and the saddle points move towards the Fermi level. We appreciate as well that the two pairs of bands immediately over and below the cones lower their energies and start to merge. Also, in Fig.\,\ref{fig:Bmesh}(d) and (e), higher weights correspond to these merging bands, whereas the cone states around the Fermi level tend to be delocalized in many $\qq$s. 
In Fig.\,\ref{fig:Bmesh}(f) the cones have collapsed into flat bands. We remark how their weight in this area of reciprocal space is small compared to that of the merging bands, despite the sharp peak on the DOS of Fig.\,\ref{fig:Bbandsdos}(c). This is an indicator of delocalization in $\qq$, and is not unexpected, as these states are known to be well localized in the AA stacking region in real space \cite{Ugeda,Felix}.

\section{\label{sec:conc}Conclusions}

We have presented a simple formulation of the band unfolding problem, a tool necessary to extract useful information from the band structure of large supercell calculations. The idea of a \textit{full unfolding} that expands the bands not only to the primitive cell, but to the full reciprocal space, allows to treat this problem as a decomposition of the wave functions into its Fourier coefficients. A \textit{refolding} recovers the conventional unfolded bands in the PBZ of the crystal. It is feasible for any eigenstate, regardless of the basis used. In the case of plane wave codes this implementation shall be almost immediate. 

We have successfully applied our algorithm to obtain new characterizations of non-trivial physical systems. The fully unfolded bands provide a distribution of the states as a function of their energy and momenta, allowing a direct comparison with experimental photoemission spectra, as well as a way to determine a value of the effective mass of the system under study in a chosen direction of reciprocal space. Refolding into the primitive cell yields clear band spectra that allow comparison with the crystal bands, even identifying crystal-like patterns in an amorphous solid. 

The outcomes of this work prove that the underlying state distribution in reciprocal space is much richer than what conventional band structures can evince, transcending the existence of any real or imposed periodicity.

\section{Akcnowledgements}
We are grateful to B. Battarai and D. Drabold for sharing the coordinates of their a-Si model with us. JMS thanks Prof. P. B. Allen for introducing him into the unfolding problem. Funded by Spain's MINECO grant FIS2015-64886-C5-5-P.

\bibliographystyle{apsrev4-1}

\begin{thebibliography}{99}

\bibitem{Boykin2005}
T. B. Boykin and G. Klimeck,
Phys. Rev. B {\bf 71}, 115215 (2005)

\bibitem{Boykin2007}
T. B. Boykin, N. Kharche, G. Klimeck and M. Korkusinski,
J. Phys.: Condens. Matter \textbf{19}, 036203 (2007)

\bibitem{Boykin2018}
T. B. Boykin, A. Ajoy,
Physica B \textbf{531}, 130-138 (2018)

\bibitem{Dargam}
T. G. Dargam, R. B. Capaz, and B. Koiller,
Phys. Rev. B {\bf 56}, 9625 (1997)

\bibitem{Lee} C. C Lee, Y. Yamada-Takamura and T. Ozaki, 
J. Phys.: Condens. Matter \textbf{25} 345501 (2013)

% --- 6 ---
\bibitem{Chen2018}
M. X. Chen and M. Weinert, 
Phys. Rev. B \textbf{98}, 245421 (2018)

\bibitem{Ku} 
W. Ku, T. Berlijn and C. C. Lee, 
Phys. Rev. Lett.\textbf{104}, 216401 (2010).

\bibitem{Popescu10}
V. Popescu and A. Zunger,
Phys. Rev. Lett. {\bf 104}, 236403 (2010)

\bibitem{Popescu12}
V. Popescu and A. Zunger,
Phys. Rev. B {\bf 85}, 085201 (2012)

\bibitem{Allen2013}
P. B. Allen, T. Berlijn, D. A. Casavant, and J. M. Soler, Phys. Rev. B {\bf 87}, 085322 (2013)

% --- 11 ---
\bibitem{Huang}
H. Huang et al., New J. Phys. \textbf{16}, 033034 (2014)

\bibitem{Medeiros}
P. V. C. Medeiros, S. Stafstr\"om, and J. Bj\"ork, 
Phys. Rev. B \textbf{89}, 041407 (2014)

\bibitem{Rubel}
O. Rubel, A. Bokhanchuk, S. J. Ahmed, and E. Assmann,
Phys. Rev. B {\bf 90}, 115202 (2014)

\bibitem{Kosugi}
T. Kosugi, H. Nishi, Y. Kato, and Y. Matsushita, 
J. Phys. Soc. Jpn. \textbf{86}, 124717 (2017)

\bibitem{notation}
Following conventional practice, we write the Bloch wave vector $\KK$ as a subscript, even though it is a continuous variable.

% --- 16 ---
\bibitem{surface}
F. J. Himpsel and D. Eastman, 
J. Vac. Sci. Technol. \textbf{16}, 1297
(1979)

\bibitem{blackphos}
C. Q. Han et al.,
Phys. Rev. B 90, 085101 (2014)

\bibitem{cost}
 In these cases, however, the vector $\KK$ such that $\KK+\GG=\kk+\gg$, will depend not only on $\kk$ but also on $\gg$. This will make the calculation of $n_{RBZ}(\kk,\epsilon)$, at given $\kk$ points, considerably more expensive.

\bibitem{Abinit}
https://www.abinit.org/

\bibitem{QE}
http://www.quantum-espresso.org/

% --- 21 ---
\bibitem{Vasp}
http://www.vasp.at/

\bibitem{Soler2002}
J. M. Soler et al., 
J. Phys.: Condens. Matter {\bf 14}, 2745 (2002)

\bibitem{PBE} J. P. Perdew, K. Burke and M. Ernzerhof, 
Phys. Rev. Lett.\textbf{77}, 3865 (1996)

\bibitem{Igram} D. Igram, B. Bhattarai, P. Biswas and D.A. Drabold, 
J. Non-Cryst. Solids \textbf{492}, 27-32 (2018)

\bibitem{Barber} H. D. Barber,  Solid-State Electron. \textbf{10}, 1039 (1967)

% 26

\bibitem{Ugeda}
M. M. Ugeda, I. Brihuega, F. Hiebel, P. Mallet, J. Y. Veuillen, J. M. Gomez-Rodriguez, and F. Yndurain, 
Phys. Rev. B \textbf{85}, 121402(R) (2012)

\bibitem{Lastra}
J. M. Garc\'ia-Lastra,
Phys. Rev. B \textbf{82}, 235418 (2010)

%\bibitem{LopezdosSantos}
%J. M. B. Lopez dos Santos, N. M. R. Peres, A. H. Castro Neto, 
%Phys. Rev. Lett. \textbf{99} 256802-4 (2007)

\bibitem{Cao}
Y. Cao, V. Fatemi, S. Fang, K. Watanabe, T. Taniguchi, E.
Kaxiras, and P. Jarillo-Herrero, Nature (London) \textbf{556}, 43 (2018)

\bibitem{Felix}
F. Yndurain, 
Phys. Rev. B \textbf{99} 045423 (2019)

\bibitem{Bezanilla}
A, Lopez-Bezanilla,
Phys. Rev. Materials \textbf{3}, 054003 (2019)

\bibitem{Carr}
S. Carr, S. Fang, P. Jarillo-Herrero and E. Kaxiras,
Phys. Rev. B \textbf{98}, 085144 (2018)

\bibitem{Yankowitz}
M. Yankowitz, S. Chen, H. Polshyn, K.Watanabe, T. Taniguchi,
D. Graf, A. F. Young and C. R. Dean,
Science \textbf{363} 1059 (2019)

\bibitem{Berland}
K. Berland and P. Hyldgaard, Phys. Rev. B 89, 035412 (2014)

\bibitem{Roman}
G. Román-Pérez and J. M. Soler, Phys. Rev. Lett. 103, 096102
(2009)

\bibitem{Laiss}
G. Trambly de Laissardière, D. Mayou, and L. Magaud, Phys.
Rev. B \textbf{86}, 125413 (2012)


\end{thebibliography}

\end{document}